\documentclass[twocolumn, aps, showpacs]{revtex4}
\usepackage{graphics}

\begin{document}

\title{Localization of a Gaussian polymer in a weak periodic surface
potential disturbed by a single defect}
\author{Andrei A. Fedorenko and Semjon Stepanow}
\affiliation{Martin-Luther-Universit\"{a}t Halle-Wittenberg,
Fachbereich Physik, D-06099\\ Halle, Germany}
 \email{fedorenko@physik.uni-halle.de}
\date{November 21, 2002}

\begin{abstract}
Using the results of the recently studied problem of adsorption of a Gaussian polymer
in a weak periodic surface potential  we study the influence of a single rod like defect
on the polymer being localized in the periodic surface potential. We have found that
the polymer will be localized at the defect under condition $u>u_c$, where $u_c$ is
 the localization threshold  in the periodic potential, for any infinitesimal
strength of the interaction with defect.  We predict that the concentration of
monomers  of the localized polymer decays exponentially as a function of the distance
to the defect and is modulated with the period of the surface potential.
\end{abstract}

\maketitle

\section{Introduction}

Adsorption of polymers at surfaces and interfaces is of large interest in
different topics of science and technology and has been investigated
extensively in recent years [\onlinecite{edwards65}-\onlinecite{fleeretal-book}]
(and citations therein). The effects of surface heterogeneities, which is of wide
interest for different applications such as pattern recognition,
technological\ and biological applications etc., have been studied in [\onlinecite%
{hone/pincus87}-\onlinecite{huang/gupta01}]. Recently, we have considered the
problem of adsorption of a Gaussian polymer in a weak periodic surface
potential \cite{preprint}. We present here the details of these calculations
and apply the method we used in \cite{preprint} to study the localization of
a Gaussian polymer in the periodic surface potential disturbed by a single
defect.

The problem of the behaviour of a Gaussian polymer in an external potential
is equivalent to the problem of the behaviour of a quantum mechanical
particle in an external potential \cite{edwards65,degennes69} .
According to this mapping our results for polymer adsorption are valid for
localization of a quantum mechanical particle in the periodic external
potential. In the case of polymer the periodic surface potential can be
realized by the surface of a microphase separated diblock copolymer melt
(see for example \cite{bates-fredrickson,binder} and references
therein), while in the quantum mechanical counterpart of the problem in
context of semiconductor superlattices \cite{herman-book} (and references
therein).

The paper is organized as follow. After brief introduction to the model and
the Green's function method in Section 2, we consider in details the problem
of adsorption of a polymer chain in a weak periodic surface potential. In
Section 3 we study the effect of a single defect on the polymer being
localized in the periodic surface potential.

\section{Adsorption of a polymer in the periodic surface potential}

\label{Sec2}

\begin{figure}[tbp]
\resizebox{0.45\textwidth}{!}{  \includegraphics{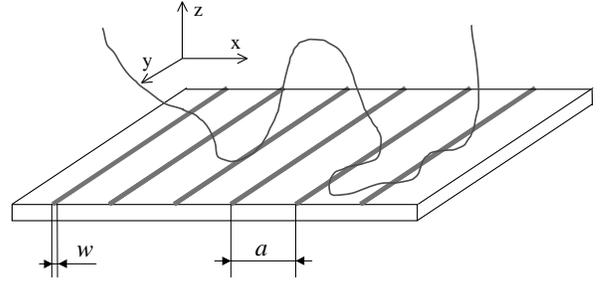}}
\caption{The periodic surface, which is modelled by the potential
in equation (1).} \label{surface}
\end{figure}

The periodic surface potential can be described by
\begin{equation}
V(x,y,z)=-u\delta (z-z_{0})\sum\limits_{n=-\infty }^{\infty }\delta (x-na),
\label{pp1}
\end{equation}
where $u$, $a$, $z_{0}$ are assumed to be positive, and $\delta (x)$ is the
Dirac's delta-function. The potential models rods which are parallel to the $%
y$ axis with the distance $a$ along the $x$ axis between the next neighbors
(see Figure \ref{surface}). The size of the rods $w$, which is suppressed in
(\ref{pp1}), is however a relevant quantity as it will be shown below. Due
to the impenetrability of the wall at $z=0$ the distance of the potential
well to the surface is chosen as $z_{0}$. As it is well-know from Quantum
Mechanics \cite{landau/lifshitzQM} the delta-function potential corresponds
to the shallow potential well. The limit to the homogeneously attracting
surface can be obtained from (\ref{pp1}) by $a\rightarrow 0$ and $%
u/a\rightarrow \mathrm{const}$ (see below).

The Green's function of a polymer in an external potential, which gives the
relative number of conformations of the chain with the ends fixed at $%
\mathbf{r} $ and $\mathbf{r}^{\prime }$, obeys the following equation
\begin{equation}
\frac{\partial }{\partial N}G(\mathbf{r},N;\mathbf{r}^{\prime })-D\nabla
^{2}G(x,z,N;x^{\prime },z^{\prime })+\frac{V(\mathbf{r})}{k_{B}T}G=0
\label{pp2}
\end{equation}
with condition $G(\mathbf{r},0;\mathbf{r}^{\prime })=\delta(\mathbf{r}-%
\mathbf{r}^{\prime })$. Here $N$ is the polymerization degree of the ideal
chain and $D=l^2/6$, where $l$ is the statistical segment length of the
chain. The equation (\ref{pp2}) can be written as an integral equation as
follows
\begin{eqnarray}
&& G(x,z,N;x^{\prime },z^{\prime }) =G_{0}(x,z,N;x^{\prime },z^{\prime })
\nonumber \\
&& -\int_{0}^{N}ds\int_{-\infty }^{\infty }dx_{1}\int_{0}^{\infty
}dz_{1}G_{0}(x,z,N-s;x_{1},z_{1})  \nonumber \\
&& \times \frac{V(x_{1},z_{1})}{k_{B}T}G(x_{1},z_{1},s;x^{\prime },z^{\prime
}),  \label{pp3}
\end{eqnarray}
where
\begin{eqnarray}
&& G_{0}(x,z,N;x^{\prime },z^{\prime })=\frac{1}{4\pi D N}\exp \left( -\frac{%
(x-x^{\prime })^{2}}{4DN}\right)  \nonumber \\
&& \times \left[ \exp \left( -\frac{(z-z^{\prime })^{2}}{4DN}\right) - \exp
\left( -\frac{(z+z^{\prime })^{2}}{4DN}\right) \right]  \label{G_0}
\end{eqnarray}
for $N\geq 0$ and is zero for $N<0$. Equation (\ref{pp2}) is related to the
Schr\"{o}dinger equation by using the replacements: $N=it$, $l^2/3k_{B}T=1/m$%
, $k_{B}T=\hbar $. The bare Green's function $G_{0}(x,z,N;x^{\prime
},z^{\prime })$ is the solution of the diffusion equation i.e. equation (\ref%
{pp2}) with $V(\mathbf{r})=0$ in the half space, $z\geq 0$, with the
Dirichlet boundary condition at $z=0$. The dependence on $y$ in equation(\ref%
{pp3}) is separated while the potential is independent of $y$. Inserting the
potential (\ref{pp1}) into (\ref{pp3}) and carrying out the Laplace
transform with respect to $N$ we arrive at
\begin{eqnarray}
&& G(x,z,p;x^{\prime },z^{\prime })=G_{0}(x,z,p;x^{\prime },z^{\prime})
\nonumber \\
&& +u\sum\limits_{n=-\infty }^{\infty
}G_{0}(x,z,p;an,z_{0})G(an,z_{0},p;x^{\prime },z^{\prime }),  \label{pp4}
\end{eqnarray}
where
\begin{eqnarray}
&&G_{0}(x,z,p;x^{\prime },z^{\prime })  \nonumber \\
&& =\frac{1}{2\pi D}(K_{0}(\sqrt{(x-x^{\prime })^{2}+(z-z^{\prime })^{2}}%
\sqrt{p/D})  \nonumber \\
&& -K_{0}(\sqrt{(x-x^{\prime })^{2}+(z+z^{\prime })^{2}}\sqrt{p/D})),
\label{pp5}
\end{eqnarray}
is the Laplace transform of (\ref{G_0}), $K_{0}(x)$ is the modified Bessel
function of the second kind. Henceforth $u$ is given in units of $k_{B}T$.
In the case of adsorption onto an interface only the first term on the
right-hand side of (\ref{pp5}) appears. Restricting the summation in (\ref%
{pp4}) to only one term gives the problem of localization onto one rod.
Neglecting the $z$-dependence in (\ref{pp4}) and using
\begin{equation}
G_{0}(x,p;x^{\prime })=1/\sqrt{4Dp}\exp (-\left| x-x^{\prime }\right| \sqrt{%
p/D})  \label{pp5-kp}
\end{equation}
instead of (\ref{pp5}) gives the Green's function formulation of the
well-known Kronig-Penney model \cite{kronig-penney}. The Kronig-Penney model
was used in \cite{sommer/blumen97} to study the behaviour of a polymer in a
striped potential.

We now will solve equations (\ref{pp4}). Inserting $x=an$, $n=0$, $\pm 1$, $%
...$ and $z=z_{0}$ into (\ref{pp4}) we obtain an infinite inhomogeneous
system of equations for $G(an,z_{0},p;x^{\prime },z^{\prime })$
\begin{eqnarray}
&& G(an,z_{0},p;x^{\prime },z^{\prime })-u\sum\limits_{m=-\infty }^{\infty
}G_{0}(an,z_{0},p;am,z_{0})  \nonumber \\
&& \times G(am,z_{0},p;x^{\prime },z^{\prime })=G_{0}(an,z_{0},p;x^{\prime
},z^{\prime }).  \label{pp6}
\end{eqnarray}
The periodicity of the potential (\ref{pp1}) along the $x$ axis, which
entails the Bloch theorem for the wave function \cite{bloch}, permits to
solve the system of equations (\ref{pp6}) by using the Fourier
transformation. Assuming that the system consists of $2M+1$ rods we consider
the discrete Fourier transform for each rod-dependent quantity $F_{n}$ as
follows $F_{n}=\sum_{k}e^{ikan}f_{k}$, where $k=\frac{2\pi }{a}\frac{j}{%
(2M+1)}$, $(j=-M,...,M)$ is the quasimomentum. Substituting $%
G(an,z_{0},p;x^{\prime },z^{\prime })=$\newline
$\sum_{k}e^{ikan}g_{k}$ and $G_{0}(an,z_{0},p;x^{\prime },z^{\prime
})=\sum_{k}e^{ikan}b_{k}$ \ into (\ref{pp6}) diagonalizes the latter (in the
limit of large $M$), so that we obtain the solution as

\begin{equation}
g_{k}=\frac{b_{k}}{1-u\sum\limits_{m=-M}^{M}e^{ikam}G_{0}(am,z_{0},p;0,z_{0})%
}.  \label{pp7}
\end{equation}
The inverse Fourier transform of (\ref{pp7})\ gives
\begin{eqnarray}
&& G(an,z_{0},p;x^{\prime },z^{\prime }) =\frac{1}{2M+1} \sum_{k} \frac1{1-u
R(k,p)}  \nonumber \\
&& \times \left[ \sum_{m=-M}^{M}e^{ika(n-m)}G_{0}(am,z_{0},p;x^{\prime},
z^{\prime })\right],  \label{pp8}
\end{eqnarray}
where we have introduced the function
\begin{eqnarray}
&& R(k,p)=G_{0}(0,z_{0},p;0,z_{0})  \nonumber \\
&& +2\sum\limits_{m=1}^{\infty }\cos(mka) G_{0}(am,z_{0},p;0,z_{0})
\label{pp22}
\end{eqnarray}
Insertion of (\ref{pp8}) into (\ref{pp4}) gives $G(x,z,p;x^{\prime},z^{%
\prime })$ as
\begin{eqnarray}
&& G(x,z,p;x^{\prime },z^{\prime }) =G_{0}(x,z,p;x^{\prime },z^{\prime })
\nonumber \\
&&+\frac{u}{2M+1}\sum_{k}\sum_{n=-M}^{M}\sum_{m=-M}^{M}\frac{\exp({ika(n-m)})%
}{1-u R(k,p)}  \nonumber \\
&& \times G_{0}(x,z,p;an,z_{0})G_{0}(am,z_{0},p;x^{\prime},z^{\prime }).
\label{pp9}
\end{eqnarray}
In the limit $M\rightarrow \infty $ the sum over $k$ should be replaced by
the integral in agreement with $\sum_{k}f(k)\rightarrow (2M+1)(a/2\pi )\int
dkf(k)$. The zeros of the denominator of (\ref{pp9}) give the main
contributions to $G(x,z,N;x^{\prime },z^{\prime })$ for large $N$. Taking
into account the latter generates the spectral expansion of the Green's
function
\begin{equation}
G(x,z,N;x^{\prime },z^{\prime })\simeq \sum_{k}e^{p_{k}N}\psi _{k}(x,z)\psi
_{k}^{\ast }(x^{\prime },z^{\prime }),  \label{se}
\end{equation}
where $\psi _{k}(x,z)$ and\ \ $p_{k}$ are eigenfunctions and eigenvalues of
\ the time-independent Schr\"{o}dinger \ equation

\begin{equation}
-D\nabla ^{2}\psi _{k}(x,z)+\frac{V(\mathbf{r})}{k_{B}T}\psi
_{k}(x,z)=-p_{k}\psi _{k}(x,z).  \label{sg}
\end{equation}
Thus, the zeros of the denominator in equation (\ref{pp9}) considered as
function of $p$ gives the energy eigenvalues $-p_{k}$. The comparison of the
inverse Laplace transform of equation (\ref{pp9}) with (\ref{se}) gives the
following expression for the wave functions

\begin{equation}
\psi _{k}(x,z)\sim \sum_{n=-\infty }^{\infty
}e^{ikan}G_{0}(x,z,p_{k};an,z_{0}).  \label{ppp1}
\end{equation}
It is easy to check that (\ref{ppp1}) fulfils the Bloch theorem. Notice that
the exact wave function $\psi _{k}(x,z)$ is given as the Laplace transform
of the bare Green's function.

We now will show how to recover from (\ref{pp9}) the case of continuously
attracting surface. The first two terms in the denominator of (\ref{pp9})
corresponds to the eigenvalue condition for a rod. If $a$ is small, the sum
can be replaced by the integral as follows $\sum_{m=1}^{\infty
}f(ma)\rightarrow a^{-1}\int_{0}^{\infty }dxf(x)$. The quantity $u/a=%
\overline{u}$ is the surface density of the strength of the potential. The
integral over $x$ with $f(x)=G_{0}(x,z_{0},p;0,z_{0})$ gives
\begin{eqnarray}
 &&\mbox{}\hspace{-10mm} \int_{0}^{\infty }dx\cos  (kx)G_{0}(x,z_{0},p;0,z_{0}) \nonumber \\
 && =1/(4\sqrt{D}\sqrt{p+Dk^{2}})  \nonumber \\
&& \times (1-\exp (-2z_{0}/\sqrt{D}\sqrt{p+Dk^{2}})).  \label{pp10}
\end{eqnarray}
The case of adsorption onto a homogeneously attracting surface will be
recovered by taking the limit $u\rightarrow 0$, $a\rightarrow 0$ and $u/a=%
\overline{u}$. As it was pointed in \cite{stepanow} for the case of
continuously interface the transverse degree of freedom are decoupled to the
in-plain degree of freedom, since the latter can be integrated out in the
definition of Green's function. The full Green's function can be obtained
from the reduced Green's function by replacing the Laplace transform
variable $p$ through $p+Dk^2$, where $k$ is the wave vector corresponding to
the Fourier transformation with respect to the in-plane coordinates. The
term $Dk^2$ is simply the energy of free motion along the surface. If we are
interested in studying the localization at the surface only and do not
consider migration of polymer along the surface we can put the quasimomentum
$k=0$. The second term in the denominator of (\ref{pp9}) disappears in the
limit $u\rightarrow 0$, so that the denominator of (\ref{pp9}) gives the
correct eigenvalue condition, $1-\frac{\overline{u}}{2\sqrt{pD}}(1-\exp
(-2z_0\sqrt{p/D}))=0$, for adsorption onto a homogeneously attracting
surface with the potential $u(z)=-\overline{u}\delta (z-z_{0})$.

Equation (\ref{pp9}) applies to the Kronig-Penney model if one neglects
there the dependences on $z$. The bare Green's function is then given by
equation (\ref{pp5-kp}). As a result the denominator in (\ref{pp9}) yields
\begin{equation}
\frac{\cos (ka)-\cosh (\sqrt{a^{2}p/D})+u/(2\sqrt{Dp})\sinh (\sqrt{a^{2}p/D})%
}{\cos (ka)-\cosh (\sqrt{a^{2}p/D})}.  \label{pp12}
\end{equation}
The numerator of (\ref{pp12}) equated to zero gives the well-known energy
eigenvalue condition for Kronig-Penney model.
\begin{figure}[tbp]
\resizebox{0.48\textwidth}{!}{  \includegraphics{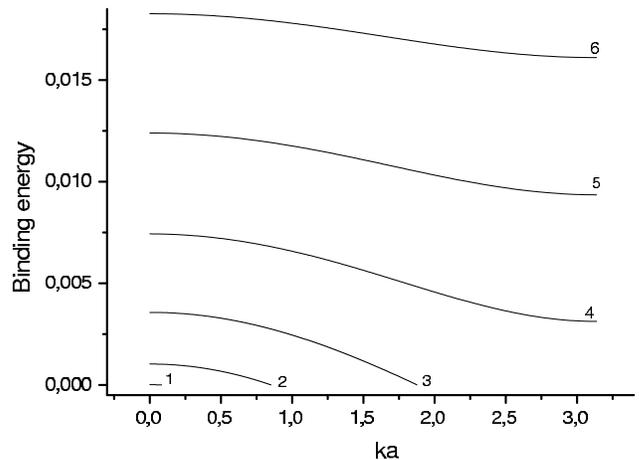}}
\caption{The localization band for $w=1$, $z_{0}=2$, $a=10$
measured in
units of $l$ for following strengths of the potential: 1. $u=0.55$; 2. $%
u=0.6 $; 3. $u=0.65$ ; 4. $u=0.7$; 5. $u=0.75$; 6. $u=0.8$. The localized
states appears for $u>u_{c}$: $u_{c}=0.54$. For $u<0.674$ the quasimomentum
of localized states does not exceed the value $k_{max}<\protect\pi /a$, for $%
u>0.674$ the localized states exist up to the edge of the Brillouin zone.}
\label{band}
\end{figure}

We now will consider the case of the periodic surface potential given by (%
\ref{pp1}). The bare Green's function obeys the Dirichlet boundary condition
at $z=0$ and is given by (\ref{G_0}). Its Laplace transform is given by
\begin{eqnarray}
&& G_{0}(am,z_{0},p;0,z_{0})=1/(2\pi D)(K_{0}(am\sqrt{p/D})  \nonumber \\
&&-K_{0}(\sqrt{a^{2}m^{2}+4z_{0}^{2}}\sqrt{p/D})).  \label{pp13}
\end{eqnarray}
The divergence of (\ref{pp13}) for $am\rightarrow 0$, which is due to
modelling the potential by Dirac's delta-function, can be avoided by
replacing $N$ in the first term of the expression $%
G_{0}(0,z_{0},N;0,z_{0})=1/(4\pi DN)-1/(4\pi DN)\exp (-z_{0}^{2}/DN)$ by $%
N+b $ with $b$ being a microscopic cutoff along the polymer (corresponds to
time in QM language ), will be related to the size of the potential well $w$%
. The necessity of introduction of the cutoff $b$ is due to the following.
The interaction potential of one rod is product of two delta functions, i.e.
the problem is two dimensional. It is well-known from Quantum mechanics \cite%
{landau/lifshitzQM} that in two dimensional potential well the eigenenergy
depends on both the depth and the width of the potential well separately.
The comparison of the binding energy in the potential $-u\delta (x)\delta
(z-z_{0})$ obtained via the present method with the energy in a shallow two
dimensional potential well \cite{landau/lifshitzQM} gives $b=w^{2}/4D$,
where $w$ is the radius of the potential well. The Laplace transform of $%
G_{0}(0,z_{0},N;0,z_{0})$ is then obtained as
\begin{equation}
G_{0}(0,z_{0},p;0,z_{0})=-\frac{e^{bp}\mathrm{Ei}(-bp)}{4\pi D}-\frac{1}{%
2\pi D}K_{0}(2\sqrt{pz_{0}^{2}/D}),  \label{pp14}
\end{equation}
where $\mathrm{Ei}(-x)=\int_{-\infty }^{-x}dt\exp (t)/t$ is the exponential
integral. Notice that we can avoid the introduction of a cutoff at
intermediate steps of the work by replacing one of delta-functions in
equation (\ref{pp1}) by the $d$-dimensional delta-function with $d<1$ and
introducing the cutoff in carrying out the limit $d\to 1$. Using (\ref{pp13}%
) and (\ref{pp14}) gives the denominator of (\ref{pp9}) as
\begin{eqnarray}
&&1+\frac{u}{4\pi D}\exp (bp)\mathrm{Ei}(-bp)+\frac{u}{2\pi D}K_{0}(2\sqrt{%
pz_{0}^{2}/D})  \nonumber \\
&&- \frac{u}{\pi D}\sum\limits_{m=1}^{\infty }\cos (amk)\left[ K_{0}(am\sqrt{%
p/D})\right.  \nonumber \\
&& \left. -K_{0}(\sqrt{a^{2}m^{2}+4z_{0}^{2}}\sqrt{p/D})\right] .
\label{pp15}
\end{eqnarray}
Considering (\ref{pp15}) as an equation for $p$ gives the spectrum of the
problem under consideration. The sum in (\ref{pp15}) cannot be performed
exactly, so that we have solved equation(\ref{pp15}) for $p$ numerically.
The localized states appears, if $u$ exceeds some threshold value $u_{c}$.
Figure \ref{band} shows the binding energy as a function of the
quasimomentum $k$ for different strengths $u$. Figure \ref{band}
demonstrates that the localized states in the periodic potential form a band
with the width depending on the strength of the potential $u$. The localized
states exist, if the quasimomentum does not exceed the value $k_{\max }$. In
contrast to a homogeneously attracting surface, the in-plane degrees of
freedom here are coupled to the transversal degree of freedom. We expect
that the width of the localization band influences the migration of the
localized polymer chain along the surface. Due to the fact that the width is
controlled by the temperature, the in-plane migration of the polymer will be
controlled by temperature. We expect that also in the quantum mechanical
counterpart of the problem the finite width band of surface localized states
will affect the in-plane properties (see below).

We now will consider the mean-square distance of one end of the polymer
chain to the surface, which has to be computed according to
\begin{eqnarray}
\langle z^2(N) \rangle = \frac{\int\limits_{-\infty}^{\infty} dx
\int\limits_{0}^{\infty} dz \int\limits_{0}^{a} dx^{\prime}z^2
G(x,z,N;x^{\prime},0)}{\int\limits_{-\infty}^{\infty} dx
\int\limits_{0}^{\infty} dz \int\limits_{0}^{a}
dx^{\prime}G(x,z,N;x^{\prime},0)} ,  \label{mean}
\end{eqnarray}
where the integration over $x^{\prime}$ is carried for simplicity over the
period of potential. To evaluate the (\ref{mean}) we have to perform the
inverse Laplace transform of (\ref{pp9}). In the limit of large chain
lengths $N\rightarrow \infty $ the main contribution to the inverse Laplace
transform appears from the residues associated with the poles. The
computation of $\ \ \langle z^{2}\rangle =\lim_{N\rightarrow \infty }\langle
z^{2}(N)\rangle $ is straightforward and results in
\begin{equation}
\langle z^{2}\rangle =\frac{2D}{p_{0}}+\frac{z_{0}^{2}}{1-\exp \left( -z_{0}%
\sqrt{p_{0}/D}\right) },
\end{equation}
where $p_{0}$ is the absolute value of the ground state eigenenergy
associated with the quasimomentum $k=0$. It is surprising that the states
with $k>0$ do not contribute to $\langle z^{2}\rangle $. We have checked
that the same holds for the mean square-distance of \ an arbitrary monomer,
and for the monomer distribution\ function $\rho (x,z)$. This is due to the
fact that after integrating over $x$ ($x^{\prime }$) in equation (\ref{pp9})
the sum over $n$ ($m$) gives $\delta (k)$ so that only the ground state
contributes to $\langle z^{2}\rangle $.
\begin{figure}[tbp]
\resizebox{0.45\textwidth}{!}{  \includegraphics{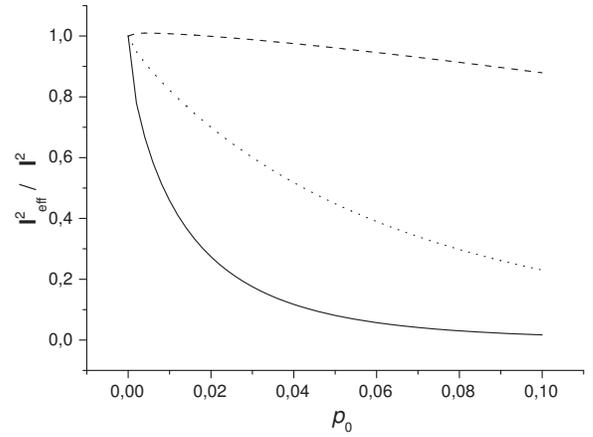}}
\caption{The effective statistical segment length as a function of
the binding energy $p_{0}$ for $w=1$, $z_{0}=2$ measured in units
of $l$. Dashed line: $a=3$; Dotted line: $a=6$; Continuous line:
$a=10$. } \label{segment}
\end{figure}

The obtained spectrum allows us to investigate the in-plane behavior of
adsorbed polymer. Due to the periodicity of the potential the long polymer
chain behaves in plane as free chain endowed with effective statistical
segment length $l^2_{\mathrm{eff}}$. To define the effective statistical
segment we compute the $x$-component of the mean-square end-to-end distance
of the adsorbed polymer for large $N$ by using equation (\ref{pp9}) and find
that
\begin{equation}  \label{l_eff}
\langle\;(x-x^{\prime })^{2}\rangle\;=\frac{1}{3}l^2_{\mathrm{eff}}N,
\end{equation}
where the effective statistical segment length
\[
l^2_{\mathrm{eff}}=-3\partial ^{2}p_{k}/\partial k^{2}\mid _{k=0}
\]
can be represented as
\[
l^2_{\mathrm{eff}}=3\left.\frac{\partial^{2}R(k,p)}{\partial k^{2}} \right|{%
\
\begin{array}{l}
\\
{}_{k=0} \\[-5pt]
{}_{p=p_{0}}%
\end{array}%
} \left[\left.\frac{\partial R(k,p)}{\partial p} \right|{\
\begin{array}{l}
\\
{}_{k=0} \\[-5pt]
{}_{p=p_{0}}%
\end{array}%
} \right]^{-1}.
\]
Notice that $l_{\mathrm{eff}}$ defined in (\ref{l_eff}) relates to the
behaviour of the polymer along the x-axes. The numerical evaluation of $l_{%
\mathrm{eff}}$ is shown in Figure \ref{segment} for three values of the
period of the potential $a$. At the localization transition $l_{\mathrm{eff}%
} $ is equal to $l$. It strongly decreases for large $p_{0}$ i.e. in the
regime of strong adsorption. For large strength of the potential well $u$
all pieces of the polymer are localized at the surface, so that in this
limit the problem of the localization in the surface periodic potential
converts to the Kronig-Penney model. It is well-known that for this model
the effective statistical segment is smaller than the bare one \cite%
{sommer/blumen97}. The squeezing of the polymer due to $l_{\mathrm{eff}}<l$
can be explained by the fact that the polymer wins energy while the portions
of the polymer make excursions along the attracting rods as it shown in
Figure \ref{surface}. This results in squeezing the polymer along the
periodicity direction. Notice that the squeezing of the polymer occurs at
the expense of the transversal size of the polymer. The size of the polymer
along the $y$-axes does not change. Figure \ref{segment} shows that the
decrease of $a$ results in a weaker decrease of $l_{\mathrm{eff}}$. The
numerical analysis of the behaviour of $l_{\mathrm{eff}}$ in the vicinity of
the localization transition for parameters $w=1$, $z_{0}=2$, $a=3$ \ yields
that $l_{\mathrm{eff}}/l=1.0044$ $>1$ for $p_{0}=0.002$, i.e. $l_{\mathrm{eff%
}}$ has a weak maximum as a function of $p_{0}$ (see Figure \ref{segment}).
The latter disappears for larger values of $a$. The condition $l_{\mathrm{eff%
}}>l$ means that the polymer stretches along the $x$-axes. Due to this the
polymer wins energy by having contacts with more rods. We expect that the
maximum is due to the rigidity of the polymer, which however cannot be
described in a more consistent way by the present model.

In context of the behaviour of a quantum particle (for example an electron)
in a surface periodic potential the motion of the particle along the surface
can be described in terms of the effective mass $m^{\ast }$, which is the
counterpart of the statistical segment length and is proportional to $l_{%
\mathrm{eff}}^{-2}$. Notice, that the above explanation of the inequality $%
l_{\mathrm{eff}}<l$  in terms of configurations of the polymer chain
implicates an explanation of the inequality $m^{\ast }>m$  in terms of
time-space trajectories of the quantum particle. The increase of the mass is
related to the size of the pieces of the trajectory localized at the same
rod. Using the well-known formula for electric conductivity in Solid State
Physics (see for example \cite{kittel}) we write the surface electric
conductivity as $\sigma =e^{2}\tau n/m^{\ast }$, where $m^{\ast }$  is the
effective mass, $e$ is the electron charge, $\tau $ is the relaxation time,
and $n$ is the surface electron density. We expect that this formula is
valid for a weakly filled band, while in the opposite case of an almost
filled band we have to take into account the effect of the delocalization of
electrons due to the external field. According to Figure \ref{segment} the
effective mass is nearly everywhere larger than the bare mass. It increases
with the increase of the strength of the potential, excepting the vicinity
of the localization transition, which is in agreement with the prediction
for Kronig-Penney model. According to the dependence of $\sigma $ on $%
m^{\ast }$ we expect that\ the surface electric conductivity is a decreasing
function of the strength of the potential. This is intuitively clear because
if the size of the pieces of the trajectory of the particle along one rod is
large, the driving field is ineffective to disengage the latter from the
rod.

\section{Localization at a single surface defect}

\label{Sec3}

The real surfaces contain various defects, so that studying their effect on
adsorption of polymers is an important question. In this section we consider
the localization of the polymer chain in a periodic surface potential
disturbed by a single defect. The surface potential in the presence of the
extended defect, which can be viewed as the additional rod situated at $%
x=x_{0}$, is modeled by the following potential
\begin{equation}
V(\mathbf{r})=-u\delta (z-z_{0})\sum\limits_{n=-\infty }^{\infty }\delta
(x-na)-v\delta (x-x_{0})\delta (z-z_{0}),  \label{pp16}
\end{equation}%
where $v$ is the strength of the defect. The method developed in Section \ref%
{Sec2} can be used in a straightforward way to study the effect of the
defect. For this aim we use the system with the periodic potential (\ref{pp1}%
) as the reference state, where the Green's function $G(x,z,N;x^{\prime
},z^{\prime })$ in the periodic potenrial is given by Eq.(\ref{pp8}).
Considering the last term in Eq.(\ref{pp16}) as perturbation we rewrite the
equation (\ref{pp2}) as an integral equation as follows
\begin{eqnarray}
&&G_{d}(x,z,N;x^{\prime },z^{\prime })=G(x,z,N;x^{\prime },z^{\prime })
\nonumber  \label{pp17} \\
&&+v\int_{0}^{N}dsG(x,z,N-s;x_{0},z_{0})G_{d}(x_{0},z_{0},s;x^{\prime
},z^{\prime }).\ \ \ \ \
\end{eqnarray}%
Henceforth the potential of defect $v$ as the strength of the periodic
potential $u$ is given in units of $k_{B}T$. The explicit expression for $%
G(x,z,N;x^{\prime },z^{\prime })$ appearing in (\ref{pp17}) can be derived
by using the inverse Laplace transform of (\ref{pp9}).
Carrying out the Laplace transform of (\ref{pp17}) we arrive at the
following algebraic equation
\begin{eqnarray}
&&G_{d}(x,z,p;x^{\prime },z^{\prime })=G(x,z,p;x^{\prime },z^{\prime })
\nonumber \\
&&+vG(x,z,p;x_{0},z_{0})G_{d}(x_{0},z_{0},p;x^{\prime },z^{\prime }).
\label{pp18}
\end{eqnarray}%
To solve this equation we substitute $x=x_{0},z=z_{0}$ and find
\begin{equation}
G_{d}(x_{0},z_{0},p;x^{\prime },z^{\prime })=\frac{G(x_{0},z_{0},p;x^{\prime
},z^{\prime })}{1-vG(x_{0},z_{0},p;x_{0},z_{0})}.  \label{pp19}
\end{equation}%
Therefore, the Green's function $G_{d}$ of the problem with potential (\ref%
{pp16}) can be expressed in terms of the Green's function $G$ of the system
with the ideal periodic potential as follows
\begin{eqnarray}
&&G_{d}(x,z,p;x^{\prime },z^{\prime })=G(x,z,p;x^{\prime },z^{\prime })
\nonumber \\
&&+v\frac{G(x,z,p;x_{0},z_{0})G(x_{0},z_{0},p;x^{\prime },z^{\prime })}{%
1-vG(x_{0},z_{0},p;x_{0},z_{0})}.  \label{pp20}
\end{eqnarray}%
The zero of the denominator in the last term of (\ref{pp20}) gives the value
of the eigenenergy of the state when polymer chain is localized at the
defect. The eigenenergy condition is obtained from (\ref{pp20}) as
\begin{eqnarray}
&&1-vG_{0}(x_{0},z_{0},p;x_{0},z_{0})-uv\frac{a}{\pi }\int\limits_{-\pi
}^{\pi }\frac{dk}{1-uR(k,p)}  \nonumber \\
&&\times \sum_{n,m=-\infty }^{\infty
}e^{ika(n-m)}G_{0}(x_{0},z_{0},p;an,z_{0})  \nonumber \\
&&\times G_{0}(am,z_{0},p;x_{0},z_{0})=0,  \label{pp21}
\end{eqnarray}%
where we have taken the limit $N\rightarrow \infty $. Equation (\ref{pp21})
cannot be solved analytically so that we analyzed it numerically. We
consider the case of extended adsorbing defect ($v>0$) which is situated at $%
x_{0}=0$. This allows us to simplify the equation (\ref{pp21}), so that we
obtain
\begin{equation}
1-vG_{0}(0,z_{0},p;0,z_{0})-uv\frac{a}{\pi }\int\limits_{-\pi }^{\pi }dk%
\frac{\left[ R(k,p)\right] ^{2}}{1-uR(k,p)}=0.  \label{pp23}
\end{equation}%
\begin{figure}[tbp]
\resizebox{0.48\textwidth}{!}{  \includegraphics{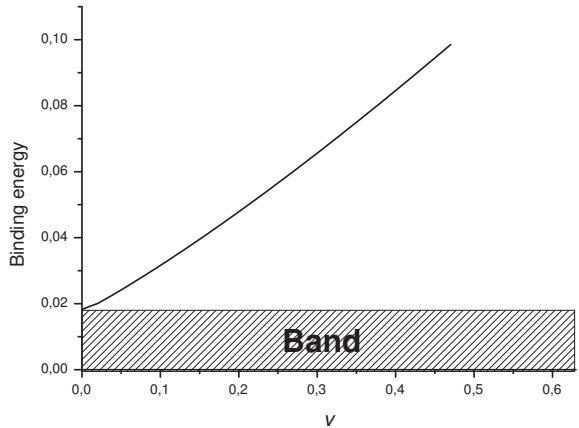}}
\caption{The binding energy of localized in-plane ground state as
function of the defect's potential $v$ for $w=1$, $a=10$,
$z_{0}=2$, $u=0.8$. At $v=0$ the energy coincide with the energy
corresponding to the upper edge of the band.} \label{defect}
\end{figure}
The numerical calculation shows that upon condition $u>u_{c}$, where $u_{c}$
is the threshold value of the periodic potential, the equation (\ref{pp23})
has a single solution $p_{d}(v)$ starting at the upper edge of the band $%
p_{d}(0)=p_{0}$ which increases with the strength of the defect $v$. The
typical dependence of $p_{d}$ on $v$ is shown in Figure \ref{defect}. Figure %
\ref{defect} shows that the polymer is localized for any infinitesimal
potential of the defect $v$, if there are localized states in the periodic
potential. The eigenfunction corresponding to the eigenvalue $p_{d}$ is
obtained from (\ref{pp20}) using the spectral expansion of the Green's
function (\ref{se}) as
\begin{eqnarray}
&&\psi _{d}\sim G_{0}(x,z,p_{d};0,z_{0})+u\frac{a}{\pi }\int\limits_{-\pi
}^{\pi }dk\frac{R(k,p_{d})}{1-uR(k,p_{d})}  \nonumber \\
&&\times \sum\limits_{m=-\infty }^{\infty }\cos
(mka)G_{0}(x,z,p_{d};am,z_{0}).  \label{pp24}
\end{eqnarray}%
\begin{figure}[tbp]
\resizebox{0.48\textwidth}{!}{  \includegraphics{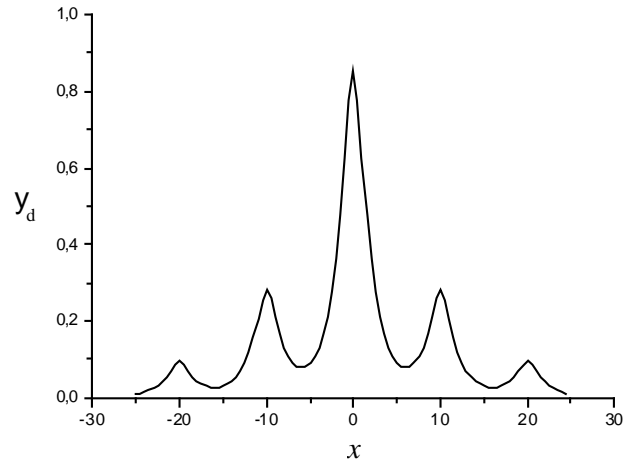}}
\caption{The profile of wave function $\protect\psi _{d}(x,z)$ in the plane $%
z=3$ for $w=1$, $a=10$, $z_{0}=2$, $u=0.8$, $v=0.01$.}
\label{wave}
\end{figure}
The function $\psi _{d}(x,z)$ decays exponentially as function of the
distance to the defect and is modulated with the period of the periodic
surface potential (see Figure \ref{wave}). Due to the gap between the
localized state and the band we have the situation of the ground state
dominance, if the polymer chain is large enough. The consequence of this is
that only the eigenfunction (\ref{pp24}) contributes to the Green's function
(\ref{se}). It is well-known that in this case the concentration of monomers
is given by $c(r)\sim |\psi _{d}(r)|^{2}$ \cite{degennes69}, so that the
oscillations of $\psi _{d}(r)$ may be observed experimentally by studying
the distribution of polymer chains on the periodic surfaces with defects.

\section{Conclusion}

We have considered the adsorption of a Gaussian polymer (and of a quantum
particle) onto an attracting surface with potential periodic along one
direction. We have found that the surface localized states form a band which
can be described by the quasimomentum entailed by the periodicity of the
surface potential. The width of the localization band depends on the
strength of the attracting potential. The binding energy decreases with
increase of the quasimomentum and becomes zero at $k=k_{\max }$, where for
not to large strengths $u$ of the periodic potential $k_{\max }$ lies within
Brillouin zone, i.e. $k_{\max }<\pi /a$. For $k_{\max }<k<\pi /a$ no
localized states exist. For sufficiently strong potential strengths (when $%
k_{\max }$ becomes equal to $\pi /a$) the polymer is always localized.

We have studied the effect of perturbation of the periodic potential by a
single rod-like defect on the adsorption of the polymer. We have found that
the defect localizes the polymer for any infinitesimal strength of the
defect potential, so that the concentration of monomers decays exponentially
with the distance to the defect and undergoes modulation associated with the
periodic surface potential. We expect, that this oscillations can be
observed experimentally by studying the distribution of polymer chains on
the periodic surfaces with defects.

The method used here can be straightforwardly applied to treat more
complicated periodic arrangements (ideal or with weak deviations from the
periodicity) of attracting wells, for example the infinite periodic set of
parallel planes along the $z$-axes with the potential in each of the plane
being periodic along the $x$- and $y$-axes.

\begin{acknowledgments}
We thank V.S.~Stepanyuk  for a useful discussion.
A support from the Deutsche Forschungsgemeinschaft (SFB 418) is
gratefully acknowledged.
\end{acknowledgments}

\end{document}